\documentclass[conference]{IEEEtran}
\IEEEoverridecommandlockouts
\usepackage{cite}
\usepackage{amsmath,amssymb,amsfonts}
\usepackage{algorithmic}
\usepackage{graphicx}
\usepackage{textcomp}
\usepackage{xcolor}
\usepackage{bm}
\usepackage{siunitx}
\robustify\bfseries
\usepackage{booktabs}
\usepackage{multirow}
\usepackage{makecell}
\usepackage[hidelinks]{hyperref}
\usepackage{pifont}
\usepackage{threeparttable}
\usepackage[capitalize,nameinlink]{cleveref}
\usepackage[ruled,vlined,linesnumbered]{algorithm2e}
\usepackage{listings}
\crefname{equation}{}{}
\Crefname{equation}{Equation}{Equations}

\usepackage{pifont}% http://ctan.org/pkg/pifont

\def\e{\ensuremath{\mathbf{e}}}

\def\x{\ensuremath{\mathbf{x}}}
\def\y{\ensuremath{\mathbf{y}}}

\newcommand\blfootnote[1]{%
  \begingroup
  \renewcommand\thefootnote{}\footnote{#1}%
  \addtocounter{footnote}{-1}%
  \endgroup
}

\def\BibTeX{{\rm B\kern-.05em{\sc i\kern-.025em b}\kern-.08em
    T\kern-.1667em\lower.7ex\hbox{E}\kern-.125emX}}
\begin{document}

% \title{Noisy Target Training for Generative \\ Speech Enhancement}
\title{Technical Report for MERL's Real-TSE \\ Challenge Submission}

\author{%
\IEEEauthorblockN{Dominik Klement$^{*1,2}$, Yoshiki Masuyama$^1$, Christoph Boeddeker$^1$, \\Kohei Saijo$^{3}$, Julius Richter$^1$, Gordon Wichern$^1$, Jonathan Le Roux$^1$}
\IEEEauthorblockA{$^1$Mitsubishi Electric Research Laboratories, Cambridge, USA, \\ $^2$Speech@FIT, Brno University of Technology, Czechia, $^3$Mitsubishi Electric Corporation, Kanagawa, Japan
}}

\maketitle

\begin{abstract}

Target speech extraction (TSE) has largely been dominated by neural network-based approaches trained and evaluated on synthetic fully overlapped data. The Real-TSE Challenge aims to advance performance on real-world far-field noisy and reverberant recordings. This technical report describes MERL's submission to the Real-TSE Challenge. Rather than proposing a novel model architecture, we built upon the baseline model and focused primarily on data preparation and cleaning. Our system was trained in four stages, beginning with pre-training on fully overlapped mixtures and simulated multi-talker conversations with noise and reverberation applied to both the mixture and the enrollment utterances. We then adapted the model to real-world conditions using noisy far-field recordings with pseudo-targets derived from processed close-talk microphone signals. Our submission achieved first place in the second track, demonstrating the critical importance of high-quality data preparation. Furthermore, we observed that DNSMOS and speaker similarity are susceptible to over-optimization, motivating an investigation of their robustness using adversarial attacks. The results show that both metrics can be driven to extreme values without degrading the token error rate or the VAD-based F1 score.
\end{abstract}


% \begin{IEEEkeywords}
% Target Speaker Extraction, Speec
% \end{IEEEkeywords}

\section{Introduction}
% \begin{itemize}
%     \item Write a short intro
%     \item Describe the model first - BSRNN + multiple conditionings, we can reference most of it, no need to be super precise there.
%     \item Describe the data pre/post processing - herem we want to describe 2 stage simulated data, what databases we used, enhancement and data filtering based on spk-sim and WER
%     \item Describe far-feld data processing - singletalker + multitalker
%     \item Describe multi-stage trainnig with referencing the results
%     \item Describe multiple objectives during fine-tuning and the tradeoff between spk-sim and TER, same for DNSMOS
%     \item Describe the attack and provide the results before/after the attack
%     \item Write conclusion.
% \end{itemize}

% Target speaker extraction (TSE) has been a long-standing problem of extracting a speaker of interest from an input speech mixture, possibly noisy or reverberant. In the past, majority of approaches has trained and evaluated on simulated data, such as Libri2Mix~\cite{...} proving substantial gains in the order of 20+dB SI-SDRi. These models, however, lack generalization to the real-world recordings, such as CHiME-6 or DipCo, which reflect how challenging the real-world audio can get. The main reason behind it is the gap between the training and the evaluation data. 
\blfootnote{$^*$Work done while Dominik Klement was an intern at MERL.}

Target speaker extraction (TSE) aims to extract a speaker of interest from a speech mixture, potentially corrupted by noise or reverberation.
Many existing approaches are trained and evaluated on simulated datasets, such as Libri2Mix~\cite{cosentino2020librimix}, where they achieve improvements exceeding 20 dB SI-SDRi.
However, these models often fail to generalize to real-world recordings, such as CHiME-6~\cite{watanabe2020chime6} and DipCo~\cite{vansegbroeck2020dipco}, which exhibit substantially more challenging acoustic conditions.
This performance degradation is primarily caused by the domain gap between simulated training data and real-world evaluation data.

% To combat this long-standing problem, the authors of Real-T dataset constructed a challenging development and evaluation set, which contains a combination of multiple English and Chinese data, post-processed to contain mixtures up to 1 minute-long conversational speech with multiple speakers.

To address this challenge, the authors of the Real-T dataset~\cite{li2025realt} introduced a challenging development and evaluation benchmark comprising both English and Chinese recordings.
The long recordings are post-processed into conversational speech mixtures of up to one minute in duration with multiple speakers.

% In this report, we present our submission to the RealTSE challenge, which builds on the baseline model architecture. Instead of focusing on developing a strong well-performing model, we instead focused most of our effort on data pre-processing and training on real far-field conversation mixtures to show that one can achieve strong results with a rather weak model but strong data processing and training pipeline.

In this report, we present our submission to the RealTSE Challenge, which builds on the baseline model architecture. 
Rather than developing a more powerful model, we focus on data preprocessing and training with real far-field conversational mixtures.
Our results demonstrate that strong performance can be achieved with a relatively simple model when paired with an effective data processing and training pipeline.

\section{Model}
We decided to use one of the provided baseline models Band-split RNN (BSRNN) with multi-level speaker conditioning~\cite{zhang2025multi} that utilizes per-frame and pooled speaker embeddings for speaker conditioning.

Compared to the baseline configuration, we replaced EcapaTDNN-512 with EcapaTDNN-1024~\cite{desplanques2020ecapa}, which is a larger and more powerful speaker embedding model. We also increased the depth of the BSRNN model from 6 to 10 blocks. This modification significantly improved the extraction performance.
% Compared to the baseline configuration, we replaced the pretrained WeSpeaker EcapaTDNN-512~\cite{} with the pretrained WeSpeaker EcapaTDNN-1024~\cite{desplanques2020ecapa}.

\section{Data Processing}
In this section, we describe our multi-stage data processing pipeline, consisting of cleaning single-speaker data for simulated pre-training and processing far-field and close-talk mixture pairs for real noisy data training.

\subsection{Pre-training Speech Data Processing}
We combine data from multiple open datasets containing vast amounts of speakers and speaking style variability: LibriSpeech~\cite{panayotov2015librispeech}, VoxCeleb2~\cite{chung2018voxceleb2}, Emilia-YODAS English and Chinese~\cite{he2025emilia}, VCTK~\cite{yamagishi2019vctk}, and EARS~\cite{richter2024ears}. 

First, we applied ClearerVoice~\cite{zhao2025clearervoice} speech enhancement models based on masking MOSSFormer~\cite{zhao2023mossformer} to preprocess the noisy corpora such as VoxCeleb2 and Emilia-YODAS. Then, we computed speaker cosine similarity between the original and the enhanced recordings using a pre-trained ResNet34 model from WeSpeaker~\cite{wang2023wespeaker} to further filter out data with speaker similarity lower than 0.85. We observed that the outliers with speaker similarity lower than 0.5 contained multiple speakers or severe degradation that the enhancement could not clean up. Furthermore, we applied another filtering to only consider audio with DNSMOS above 3.0. Even though a high DNSMOS score does not guarantee clean speech, this filtering step ensured that the enhancement likely produced estimates accurate enough for training.

Afterwards, we computed word-level alignments using Montreal Forced Aligner (MFA)~\cite{mcauliffe2017mfa} to avoid sampling non-speech segments as target speaker enrollments.
Although such segments could help train the model not to always expect the target speaker to be present in the mixture, they would make it difficult to precisely control their proportion during training. Instead, with a probability of 5\%, we explicitly generated non-target samples by pairing the enrollment with a speaker absent from the input mixture and trained the model to produce a zero output.

\subsection{Noises and Room Impulse Responses}
We applied room impulse responses (RIRs) on a clean input mixture and enrollment independently with a probability of 0.8. The RIRs were randomly sampled from DNS4~\cite{dubey2022dnschallenge} and from synthetic impulse responses simulated using Pyroomacoustics~\cite{scheibler2018pyroomacoustics} with weights 0.8 and 0.2, respectively. We purposely simulated larger rooms to cover the scenarios of extreme echoes.

Furthermore, to increase the robustness against multiple noises, we used the following noise databases containing more stationary and also impulsive noises: CHiME-3~\cite{barker2015chime3}, DEMAND~\cite{thiemann2013demand}, DNS4~\cite{dubey2022dnschallenge}, FMA~\cite{defferrard2017fma}, FSD50K~\cite{fonseca2022fsd50k}, MUSAN~\cite{snyder2015musan}, WHAM!~\cite{Wichern2019-gp}, and simulated Wind noises from the URGENT challenge~\cite{zhang2024urgent}.

We used weighted sampling to increase the weight towards the stationary-like noises, where we set the probability of sampling CHiME3, WHAM!, and Wind noises to 0.15, and the probability of all the other noises to 0.11.
We randomly sampled SNR from a $[-5, 15]$~dB range and added the noise to the input mixture and enrollment with probability 0.8. Mixture and enrollment noises were selected independently, covering all combinations of clean and noisy mixture and enrollment pairs during training.

\subsection{Real Data Processing}
To include real-world data for training, we utilized CHiME-6~\cite{watanabe2020chime6}, AMI~\cite{carletta2005ami}, AISHELL-4~\cite{fu2021aishell4}, and AISHELL-5~\cite{dai2025aishell5}. As AISHELL-5 provides close-talk recordings that are already synchronized with far-field microphones and do not contain significant cross-talk, we only perform speech enhancement using ClearerVoice. For AISHELL-4, we perform GSS~\cite{boeddecker2018gss} on multi-channel audio using ground-truth speaker diarization and apply speech enhancement afterwards. During training, we use both AISHELL datasets for simulated far-field mixtures only, as both contain a low amount of overlapped speech.

\subsubsection*{Single-talker Far-field Mixtures}
To further adapt the model to real noisy data, we select single-talker segments of speech based on the provided dataset annotations. Then, we cut out the same segment from the speaker's close-talk microphone and average all the close-talk channels if multiple are available, and proceed with applying ClearerVoice speech enhancement to denoise and dereverberate the close talk. 

To enable training on such mixtures, we perform max-peak cross-correlation synchronization with maximum lag set to one second.
Afterwards, inspired by~\cite{nakatani2026generating}, we apply a causal Wiener filter with 10~ms window (160 taps) to project the close-talk signal to the corresponding far-field. To ensure that such filter can be estimated, we shift the close-talk microphone by a quarter of the width of the Wiener filter such that the far-field recording is ahead of the corresponding close-talk.

After projecting the close-talk to the corresponding far-field, we observed that the projected signal contains noises, caused by the fact that the far-field recording contains noise. 
Hence, we apply speech enhancement once more at the very end to clean up the projected close-talk.

Before training, we use Nvidia Parakeet-V2~\cite{sekoyan2025canary} to filter out all the utterances with WER higher than 30\%. We do not perform WER-based filtering on AISHELL datasets, as these are relatively simple and clean data compared to CHiME-6 or AMI.

During mixing, to prevent potential device or channel classification, we simulate far-field mixtures on-the-fly using the speakers from the same session, device, and the same channel.

\subsubsection*{Real Far-field Mixture Training}
First, we find multi-talker conversational segments with max duration 30~s with at least 2 active speakers. Then, we find an enrollment for each speaker that is at least 10 seconds long, the majority of which is speech. Afterwards, we use the best-performing TSE model trained on far-field mixtures to reduce cross-talk on CHiME-6 and AMI close-talk microphones. The Parakeet-V2 ASR is then used to filter out segments with WER above 50\% on CHiME-6 and 30\% on AMI, keeping the rest of the close-talk far-field pairs for training. At the end, we perform synchronization and close-to-distant projection as described above, and apply speech enhancement to get rid of the projection noise.

\section{Training Curriculum}
It has been shown in~\cite{bengio2009curriculum} that multi-stage curriculum pre-training starting from easier data and gradually progressing towards more challenging data increases model performance, generalization, and convergence. Therefore, we performed training in 4-stages.

\subsection{Fully-overlapped Pre-training}
We start with pre-training the TSE model from scratch using fully-overlapped Libri2Mix-style data simulated on-the-fly with randomly sampled enrollments.

During this stage, the model should learn the mapping between the enrollment and the mixture, and extract the target speaker. This training phase is performed for 200k training steps.

\subsection{Noisy Simulated Conversations}
Next, we initialize the model from the previous stage and train the model using on-the-fly generated conversational mixtures simulated by a modified FastMSS~\cite{polok2026mind} simulator.
During this stage, we also introduce reverberation and noise to increase the model's robustness.

\subsection{Far-field Mixtures}
After the second pre-training stage, we continue training on on-the-fly simulated far-field mixtures generated from single-talker segments of CHiME-6, AMI, AISHELL-4, and AISHELL-5. Enrollment utterances are randomly sampled from the same segments. Since AMI does not contain background noise, we augment both the mixtures and the enrollment utterances with artificial noise with a probability of 0.2.

\subsection{Real Far-field Mixtures}
Lastly, the fourth stage contains real noisy mixture training against targets that were extracted by our previous TSE model. To increase the variability during this training stage, we mix in 20\% of synthetic data, and around 30\% of far-field mixtures from the third stage of training. 

\section{Training Setup}
During the first two stages, we optimized the following time-domain loss function:
\begin{equation}
    \label{eq:log_mae_loss}
    \mathcal{L}^\text{time} = \log \left( \frac{1}{B} \sum_{i=1}^B |\y_i - \hat{\x}_i| \right),
\end{equation}
where $\hat{\x}_i$ is the prediction and $\y_i$ is the corresponding target, and $B$ is the mini-batch size. Optimizing this loss instead of SI-SNR or SNR prevents numerical errors for zero targets (i.e., total silence). The loss function was optimized using AdamW~\cite{loshchilov2019decoupled} optimizer with learning rate (LR) $0.0005$ and weight decay $0.001$.
We used the ReduceLROnPlateau\footnote{\href{https://docs.pytorch.org/docs/2.12/generated/torch.optim.lr\_scheduler.ReduceLROnPlateau.html}{PyTorch - ReduceLROnPlateau}} learning rate scheduler according to the validation SNR.
The training recordings were randomly cropped to eight seconds while target speaker speech presence was ensured according to the aligned word-level timestamps.

During the last two stages, we used linear LR warmup for 2k steps with peak LR set to $1e-4$ and then applied cosine learning rate scheduler~\cite{loshchilov2017sgdr}.
Because the targets are enhanced projections and not the true clean signals, the phase between the far-field mixture and the projected close-talk might not be well aligned. Hence, we utilize magnitude-based multi-scale STFT (MS-STFT) loss during the last two training stages, defined as:
\begin{align}
    \label{eq:ms_stft}
    \mathcal{L}^\text{stft} = \sum_{i=1}^B \sum_{k=1}^K
    &\left\lVert |S(\y_i, w_k, h_k)|  - |S(\hat{\x}_i, w_k, h_k)| \right\rVert_1,
\end{align}
where $S(\y_i, w_k, h_k)$ represents STFT computed with the window length $w_k$ and hop length $h_k$. We selected $h_k=\frac{1}{4}w_k$ samples. We used 5 window lengths: $100, 200, 400, 800, 1600$, which corresponds to $6.25$~ms, $12.5$~ms, $25$~ms, $50$~ms, $100$~ms.
As described before, during the fourth stage, we mixed in a small portion of simulated data, so that we could utilize the time-domain loss $\mathcal{L}^\text{time}$ on that data.

Besides the reconstruction losses, we also utilized metric-aware losses. First, we used a torch-based implementation of DNSMOS that allowed us to backpropagate through it and optimize the following loss function:
\begin{equation}
    \label{eq:dnsmos_loss}
    \mathcal{L}^\text{MOS} = \frac{1}{B} \sum_{i=1}^B \left( 5.0 - \text{DNSMOS}_\text{ovrl}(\hat{\x}_i) \right)^2.
\end{equation}

Similarly, to slightly push the model to better preserve the enrollment speaker identity, we optimized a cosine similarity between the enrollment and the output:
\begin{equation}
    \label{eq:spk_sim_loss}
    \mathcal{L}^\text{spk} = \frac{1}{B} \sum_{i=1}^B \left| 1.0 - \cos(\e^\text{enrl}_i, \e^\text{pred}_i) \right|,
\end{equation}
where $\e^\text{enrl}_i, \e^\text{pred}_i$ represent the corresponding speaker embeddings extracted using ResNet34 from the enrollment and the prediction, respectively.

The final loss for the third and fourth stage training is a weighted combination of all the preceeding losses:
\begin{equation}
    \label{eq:overall_loss}
    \mathcal{L} = 10 \mathcal{L}^\text{stft} + \mathcal{L}^\text{time} + 0.01 \mathcal{L}^\text{MOS} + \mathcal{L}^\text{spk},
\end{equation}
where $\mathcal{L}^\text{time}$ is only computed on the simulated part of the training batch.

\section{Results}

% \begin{table}[h]
% \centering
% \caption{Comparison between different training stages on the DEV Set. The last ``Pure'' model was trained only using reconstruction losses.}
% \begin{tabular}{lcccc}
% \toprule
% Model & TER & F1 & SPK-SIM & DNSMOS \\
% \midrule
% 1st stage & 0.53 & 0.86 & 0.48 & 2.42 \\
% + 2nd stage & 0.44 & 0.86 & 0.47 & 2.37 \\
% + 3rd stage & 0.39 & \textbf{0.88} & 0.54 & 2.81 \\
% + 4th stage & \textbf{0.37} & \textbf{0.88} & \textbf{0.55} & \textbf{2.84} \\
% \midrule
% 4 stage, Pure & 0.37 & 0.87 & 0.47 & 2.89 \\
% \bottomrule
% \end{tabular}
% \label{tab:stage_comparison}
% \end{table}

\begin{table}[t!]
\centering
\caption{Comparison between the challenge baseline, different training stages and the adversarial attack on the DEV Set.. P stands for precision, R for recall, and SPK for speaker to enrollment cosine similarity. The model in the last row marked with $^{\star}$ was trained only using reconstruction losses during all stages.}
\begin{tabular}{lccccccc}
\toprule
% \multirow{2}{*}{Model} & \multirow{2}{*}{TER} & \multirow{2}{*}{P} & \multirow{2}{*}{R} & \multirow{2}{*}{F1} & \multirow{2}{*}{SPK} & \multicolumn{2}{c}{DNSMOS} \\
% \cmidrule(lr){7-8}
% & & & & & & OVRL & P.808 \\
& & & & & & \multicolumn{2}{c}{DNSMOS} \\
\cmidrule(lr){7-8}
Model & TER & P & R & F1 & SPK & OVRL & P.808 \\
\midrule
1st stage & 0.53 & 0.82 & 0.93 & 0.86 & 0.48 & 2.42 & 3.21 \\
+ 2nd stage & 0.44 & 0.88 & 0.88 & 0.86 & 0.47 & 2.37 & 3.15 \\
+ 3rd stage & 0.39 & 0.91 & 0.86 & \textbf{0.88} & 0.54 & 2.81 & 3.15 \\
+ 4th stage & \textbf{0.37} & \textbf{0.93} & 0.85 & \textbf{0.88} & 0.55 & \textbf{2.84} & \textbf{3.33} \\
% + attack & \textbf{0.37} & \textbf{0.93} & 0.85 & \textbf{0.88} & \textbf{0.99} & \textbf{4.07} & \textbf{3.33} \\

\midrule
\textcolor{black!50}{4 stages, Pure} & \textcolor{black!50}{0.37} & \textcolor{black!50}{0.93} & \textcolor{black!50}{0.84} & \textcolor{black!50}{0.87} & \textcolor{black!50}{0.47} & \textcolor{black!50}{2.89} & \textcolor{black!50}{3.32} \\
\bottomrule
\end{tabular}
\label{tab:stage_comparison}
% \vspace{-10pt}
\end{table}

Table~\ref{tab:stage_comparison} shows the progressive improvements in all of the metrics. First, TER significantly improves between the first and the second stage, proving that conversational simulated data with heavy augmentations improve the TSE performance on challenging far-field data. Furthermore, training with 3rd and 4th stage shows that using real noisy and reverberated data for training improves the performance even further.

In terms of spk-sim and DNSMOS, the main improvement comes from slight metric tuning, which we used to balance out these metrics with TER and F1. However, we observed that training using these metrics did not improve the perceptual quality when listening to the extracted audio. We even observed that over-tuning to speaker similarity causes the TSE model to output target-speaker-like artifacts in the presence of noise even if no one speaks. This further proves how brittle the speaker similarity models are.

To show the effects metric optimization had on our submission, we performed a four-stage training again without using the metric-aware losses during the last two stages. In this case, the model was trained using $\mathcal{L}^\text{stft}$ on all data and $\mathcal{L}^\text{time}$ on simulated-only data. The last row (``Pure'') in Table~\ref{tab:stage_comparison} shows that we could achieve the same TER and DNSMOS as with metric-aware fine-tuning. Also, $F1$ score only slightly decreased compared to the metric-aware training model due to the lower recall. This was likely caused by not optimizing for spk-sim, which usually forced the model to leak more noise that could potentially trigger the VAD system. The only metric that got significantly decreased is speaker similarity, whose value remains equal to that at the end of the second training stage. After listening to a few examples between the metric-optimized and the pure model, we could not hear any significant difference in terms of extracted speaker identity, proving how overoptimistic the metric value can get once we start optimizing it.

% However, we were able to achieve similar results comparable to the 4th stage in Table~\ref{tab:stage_comparison} without any metric-aware training whatsoever except fofr speaker-similarity, proving that DNSMOS of 2.7-2.8 can be achieved even without explcitly optimizing it.

% \todo{Infer and add such result! - WIP}

% \todo{Add per-metric (DNSMOS, SPKSIM, BOTH) + all metric optimization comparison.}

% \todo{Add motivation for the attack - clean data DNSMOS is aroun 3.2-3.3, but some systems already exceeded it, suggesting some metric-over-optimization.}

% \todo{Motivate with Godhart's law - optimization on metrics = metric is useless for evaluation.}

% \todo{Refernce the first attack paper.}

% \todo{We want to say ot loud that we did not inted to cheat!}

\section{Metric Attack}
% \begin{quote}
% When a measure becomes a target, it ceases to be a good measure. (Goodhart’s Law)
% \end{quote}
\begin{quote}
\emph{``When a measure becomes a target, it ceases to be a good measure.''}

\hfill --- Goodhart's law
\end{quote}

Many speech separation and target speech extraction systems optimize evaluation metrics either explicitly or implicitly during training, including metrics such as SI-SDR and PESQ. From the perspective of Goodhart's Law, this practice fundamentally compromises the validity of such metrics as evaluation tools: once a metric becomes an optimization target, it ceases to function as an independent measure of system quality.

This issue is particularly evident for non-intrusive neural-network-based metrics. For example, clean speech recordings from datasets such as VCTK achieve DNSMOS scores of only around 3.2~\cite{Jung2025-yc}. Therefore, observing multiple leaderboard submissions with DNSMOS scores exceeding 3.5 raised concerns about the susceptibility of model-based metrics to over-optimization. 

Neural networks are universal function approximators that can approximate arbitrary functions given sufficient capacity and width~\cite{HORNIK1989359}. Consequently, when a metric is directly optimized during training, a sufficiently expressive model can effectively learn the metric itself and, in the extreme case, internalize at training time an adversarial attack required to maximize its score. The main technical challenge to do so is carefully balancing the optimization process to avoid degrading performance on other evaluation criteria such as TER and F1. Given the better-than-clean-speech DNSMOS scores we observed on the leaderboard, we suspect some teams may have unintentionally performed such an adversarial attack.

To demonstrate the upper bound of metric-aware training and illustrate how metric optimization during training can invalidate the optimized metrics as evaluation measures, we conducted a series of per-utterance adversarial attacks using different metric combinations. We were surprised by how easily a barely perceptible perturbation could be added to the target speech extraction output waveform, yielding state-of-the-art (but perceptually meaningless) DNSMOS and speaker-similarity scores while preserving performance on the remaining metrics, including TER and F1.

\begin{algorithm}[t]
\caption{Metric Attack.}
\label{alg:attack}
\begin{lstlisting}[language=Python, mathescape=true, escapeinside={(*}{*)}, basicstyle=\footnotesize\ttfamily, aboveskip=2pt, belowskip=2pt, lineskip=-0.2ex]
def attack(inp, enrl):
  delta = nn.Parameter(torch.zeros_like(inp))
  optim = Adam([delta], lr=1e-3)

  for i in range(200):
    delta_t = tanh(delta) * 0.01
    audio_t = (inp + delta_t).clamp(-1, 1)
    loss = (stft(audio_t).abs() - stft(inp).abs())
    loss = loss.pow().mean()
    loss = loss + (5.0 - dnsmos(inp))^2
    loss = loss + (1.0 - spksim(inp, enrl))

    optim.zero_grads()
    loss.backward()
    optim.step()
  
  return audio_t  
\end{lstlisting}
\end{algorithm}

We based our attack algorithm on~\cite{liao2026beyond}, which is straightforward and described in Algorithm~\ref{alg:attack}. It performs 200 steps, in which we first calculate a small delta and add it to the waveform, then compute all the losses and sum them with unit weights. It is important to note that we heavily regularize the adversarial noise energy and also minimize the difference between the original and modified magnitude spectrograms to prevent large changes. Finally, we backpropagate the loss gradients through the metric models and update the delta using the Adam optimizer. We observed during the experiments that no special loss weights, learning rate, or number of steps tuning was required for successful convergence, proving the simplicity of the attack.

% \begin{table}[h]
% \centering
% \caption{Comparison between before and after metric attack.}
% \begin{tabular}{lcccc}
% \toprule
% Attack & TER & F1 & SPK & DNSMOS \\
% \midrule
% None & \textbf{0.37} & \textbf{0.88} & 0.52 & 2.83 \\
% DNSMOS & \textbf{0.37} & \textbf{0.88} & 0.52 & \textbf{4.18} \\
% SPK-SIM (EN) & \textbf{0.37} & \textbf{0.88} & 0.82 & 2.80 \\
% ~~+ DNSMOS & \textbf{0.37} & \textbf{0.88} & 0.82 & 4.11 \\
% ~~~~+ SPK-SIM (CN) & \textbf{0.37} & \textbf{0.88} & \textbf{0.99} & 4.07 \\
% \bottomrule
% \end{tabular}
% \label{tab:attack_comparison}
% \end{table}

\begin{table}[h]
\centering
\caption{Comparison between before and after metric attack.}
% \colsep
\setlength{\tabcolsep}{4.6pt} % default is 6pt
\begin{tabular}{lcccccc@{~~}c}
\toprule
\multirow{2}{*}{Attack} & \multirow{2}{*}{TER} & \multirow{2}{*}{P} & \multirow{2}{*}{R} & \multirow{2}{*}{F1} & \multirow{2}{*}{SPK} & \multicolumn{2}{c}{DNSMOS} \\
\cmidrule(lr){7-8}
 &  &  &  &  &  & OVRL & P808 \\
\midrule
None & \textbf{0.37} & \textbf{0.94} & \textbf{0.85} & \textbf{0.88} & 0.52 & 2.83 & \textbf{3.33} \\
DNSMOS & \textbf{0.37} & \textbf{0.94} & \textbf{0.85} & \textbf{0.88} & 0.52 & \textbf{4.18} & \textbf{3.33} \\
SPK-SIM (EN) & \textbf{0.37} & \textbf{0.94} & \textbf{0.85} & \textbf{0.88} & 0.82 & 2.80 & \textbf{3.33} \\
~~+ DNSMOS & \textbf{0.37} & \textbf{0.94} & \textbf{0.85} & \textbf{0.88} & 0.82 & 4.11 & \textbf{3.33} \\
~~~~+ SPK-SIM (CN) & \textbf{0.37} & \textbf{0.94} & \textbf{0.85} & \textbf{0.88} & \textbf{0.99} & 4.07 & \textbf{3.33} \\
\bottomrule
\end{tabular}
\label{tab:attack_comparison}
\end{table}

Table~\ref{tab:attack_comparison} shows the model results before the adversarial attack in the first row. It can be seen that attacking DNSMOS only (2nd row) does not change the unattacked metrics. Also, we can see that attacking only the English spk-sim model does not achieve an overall score close to 1, as the Chinese speaker similarity stays at around 0.5 even after the attack. The last row shows the scenario where DNSMOS and speaker models for English and Chinese are attacked. Once again, we can see that TER and F1 stayed the same, while spk-sim and DNSMOS are achieving their extremes. It is noteworthy to observe that attacking only DNSMOS achieves the best DNSMOS score out of all the attacks. The score decreases as we increase the number of simultaneously-attacked metrics. However, the relative change in DNSMOS between attacking all the metrics and only DNSMOS is negligible relative to its value.

Given the fragility of non-intrusive metrics as demonstrated by our attack and also shown in~\cite{liao2026beyond, huang2026attackingutmosprobingrobustness}, we suggest that the Challenge Organizers either remove DNSMOS and spk-sim when calculating the official ranking or replace them with alternative speech quality and speaker similarity metrics that were not attacked either advertently or inadvertently by submitted systems.

\section{Conclusion}
In this work, we described our submission to the RealTSE challenge, consisting of the baseline TSE model architecture and multi-stage data processing and training curriculum. Our results confirmed that each training stage improved the metrics, proving that thorough data preparation with heavy mixture and enrollment augmentations and real noisy data training can push the performance even further. We also showed that metrics such as speaker similarity or DNSMOS are brittle and susceptible to adversarial attacks. To push it to the extreme, we applied a per-waveform adversarial attack, which serves as an upperbound of what can be potentially achieved with training-time optimization if enough capacity and training balance is achieved. This raises challenges for the organization of future challenges in an age where generative models are thriving, and suggests the need for further metric development, which will be focus of our future work.

% \newpage\

\bibliographystyle{IEEEtran}
\bibliography{refs}

\end{document}